\documentstyle[12pt]{article}

\begin{document}

\baselineskip 0.8cm

\setlength{\textwidth}{14cm}
\setlength{\textheight}{21cm}
\setlength{\topmargin}{0cm}
\setlength{\parindent}{1cm}
\newcommand{\gsim}
{ \mbox{ \raisebox{-1ex}{ $\stackrel{>}{\sim}$ } } }
\newcommand{\lsim}
{ \mbox{ \raisebox{-1ex}{ $\stackrel{<}{\sim}$ } } }

\begin{titlepage}

\begin{flushright}
TU-410 \\
IFP-436-UNC \\
August, 1992
\end{flushright}

\vskip 1.0cm

\begin{center}
{\Large \bf Phenomenological Aspects of \\ Supersymmetric  Standard Models \\
 without Grand Unification } \\
\vskip 1.0cm

Satoshi Mizuta$^{a)}$, Daniel Ng$^{b)}$ \footnote{Address after September 1,
1992: TRIUMF, 4004 Wesbrook Mall, Vancouver, B.C., V6T 2A3, Canada}
 and Masahiro Yamaguchi$^{c)}$

\vskip 0.8cm

{\it a)  Department of Physics, Tohoku University,
Sendai 980, Japan}

\vskip 0.5cm

{\it b) Institute of Field Physics,
Department of Physics and Astronomy,
University of North Carolina,
Chapel Hill, NC 27599-3255, USA}

\vskip 0.5cm

{\it c) Department of Physics, College of General Education  \\
Tohoku University, Sendai 980, Japan}
\end{center}

\abstract{When the GUT relation on  gaugino masses is relaxed,
 the mass and composition of the lightest neutralino are different
from those in the GUT case. We discuss its
phenomenological implications on  the relic abundance of the neutralinos
and on superparticle searches. In particular, we focus on the case where the
neutral component of Winos dominates the lightest neutralino.
It turns out the Wino-LSP is not a  candidate for the dark matter.}
\end{titlepage}


In supersymmetric (SUSY) $SU(5)$ grand-unified theories \cite{SU5GUT}
and many other SUSY
GUTs,
a  one-loop renormalization group (RG) \cite{Inoue}
analysis gives us the following relation for the gaugino
mass parameters
\begin{equation}
  M_1:M_2:M_3=5/3g_1^2 : g_2^2 : g_3^2 \simeq 1:2:7, \label{GUTrel}
\end{equation}
where $g_i$ ($i=$1, 2, 3)
are the gauge coupling constants of the gauge groups
$U(1)_Y$, $SU(2)_L$ and $SU(3)_C$, respectively, and $M_i$ ($i=$1, 2, 3)
the corresponding gaugino mass parameters. Note that this relation holds,
independent of the particle content between the electroweak scale and
the GUT scale. Therefore, eq.~(\ref{GUTrel}) provides an important test
of the idea of the GUTs.

There are, however, many well-motivated models in which the relation (1) is
not satisfied. Examples include $SU(3)_C \times SU(2)_L
\times U(1)_Y$ string models \cite{string} and a flipped $SU(5)$ model
\cite{Antoniadis}
whose gauge structure
is $SU(5) \times U(1)_X$. In the former, $M_1$, $M_2$ and $M_3$
are unknown parameters which will be determined by string dynamics,
whereas in the latter, $M_1$ is independent
of the others. Indeed  in the flipped $SU(5)$ it has been argued \cite{Drees}
that, using  a RG analysis,  $|M_1| \gg
|M_2|$ is favored to achieve the correct gauge symmetry breaking
$SU(5) \times U(1)_X \rightarrow SU(3)_C \times SU(2)_L \times U(1)_Y
\rightarrow SU(3)_C \times U(1)_{em}$.

In this paper, we will consider phenomenological aspects of SUSY standard
models without the GUT relation (\ref{GUTrel}) which is assumed in  most of the
literature (see however refs. \cite{DreesTata,GriestRoszkowski,Yamaguchi}).
In particular, we are concerned with the mass ratio $r \equiv M_1 /M_2$,
since it is an important parameter in the
neutralino mass matrix as we will explain later.
We assume, as usual, that  the lightest neutralino is
also the lightest  superparticle (LSP). To clarify its properties is thus
important in SUSY phenomenology
from the viewpoints of both accelerator physics and cosmology,
especially when the LSP is stable due to a $Z_2$ symmetry called
$R$-parity.
In this paper we will mainly discuss the case of $|r|>1$, in which
the neutral component of  Winos can be a dominant composition  of the LSP
(see below).
Note that this Wino-dominant LSP is never  realized under the GUT assumption.
On the other hand, some of the details on the case of $|r| \ll 1$, where
the LSP is mostly the Bino, have been discussed in \cite{GriestRoszkowski}.

Throughout this paper, we consider the case of the minimal particle content
in the SUSY standard model. With the $R$-parity
conservation, the LSP, being a linear combination of
four neutralinos (i.e. a Bino $\tilde B$, a neutral Wino $\tilde W_3$
and two neutral Higgsinos $\tilde H_1$, $\tilde H_2$)
is stable and can  be a candidate for the dark matter \cite{darkmatter}.

 In the following, we will first investigate
the mass  and composition of the lightest  neutralino. We will find that it
is very degenerate with  the lightest chargino at the
tree-level for $|r|>1$, in particular when the lightest neutralino is
dominated by the neutral Wino.
We will examine if the neutralino is indeed lighter than the
chargino after including  radiative corrections.  We will then discuss a
cosmological implication.
It  will turn out when $|r|>1$ the relic abundance of the
neutralino LSPs  is  small in most of the
parameter space and the LSP is not a candidate for the dark matter  of the
Universe. Finally we will discuss superparticle searches in  the non-GUT case.


The mass matrix  of the neutralinos is parameterized by the gaugino mass
parameters $M_2$, $M_1(=rM_2)$, a Higgsino mass parameter $\mu$ and
the ratio of the vacuum expectation values of the two neutral Higgs bosons
$\tan \beta = v_2/v_1$
\cite{GunionHaber}. It is explicitly given by
\begin{equation}
\pmatrix{M_1 & 0 & -m_Z \sin \theta_W \cos \beta
                       & m_Z \sin \theta_W \sin \beta
\cr
         0 & M_2 & m_Z \cos \theta_W \cos \beta
                       & -m_Z \cos \theta_W \sin \beta
\cr
         -m_Z \sin \theta_W \cos \beta & m_Z \cos \theta_W \cos \beta
                           & 0 & -\mu
\cr
         m_Z \sin \theta_W \sin \beta & -m_Z \cos \theta_W \sin \beta
                           & -\mu & 0\cr} \; ,
\end{equation}
in the $(\tilde B, \tilde W_3, \tilde H_1, \tilde H_2)^T$ basis. Here $M_1$,
$M_2$ and $\mu$ are assumed to be real.
As a convention, we take $M_2>0$ and $\mu$ both positive and negative.
It is straightforward to diagonalize the matrix  and we  will denote the masses
(which are taken to be positive) by $m_{\chi^0_i}$ ($i=$1,$\cdots$,4) with
$m_{\chi^0_1} < \cdots <m_{\chi^0_4}$. The lightest neutralino $\chi_1^0$ is
written
\begin{eqnarray}
    & &  \chi_1^0  =  Z_{11} \tilde B +Z_{12} \tilde W_3
                  +Z_{13} \tilde H_1 + Z_{14} \tilde H_2, \cr
    & & \sum_{i=1}^{4 }Z_{1i}^{\ 2}  = 1
\end{eqnarray}
where $Z_{1i}$ are real numbers. It is convenient to define the purity
of Bino and Wino component,
\begin{equation}
    p(\tilde B) = Z_{11}^{\ 2}
\end{equation}
and
\begin{equation}
      p(\tilde W) = Z_{12}^{\ 2},
\end{equation}
respectively.
In fig.~1, we plot in the $(M_2, \mu)$ plane the mass of the lightest
neutralino, $m_{\chi_1^0}$, for (a) $r=0.1$ (b) $r=0.5$ (the GUT case) (c)
$r=5$.  Here we take $\tan \beta =2$.
To demonstrate the composition of $\chi_1^0$, we also plot in the same figures
the Bino purity $p(\tilde B)$ for the  cases (a), (b) and the Wino purity
$p(\tilde W)$ for the case (c).
We find as expected that for small $ r$, a gaugino-like LSP is indeed
Bino-dominant,  whereas
for  $|r|>1$, the lightest neutralino can  be dominated
by the neutral Wino
when $M_2 \ll |\mu|$.
A neutralino with a smaller mass is realized for a smaller value of $|r|$ with
$M_2$ and $\mu$ fixed.
 We will study
the mass difference between  $\chi_1^0$ and the lightest chargino
$\chi_1^{\pm}$ later.
In fig. 1, we have  included the constraints from the LEP experiments
\cite{LEP}:

(i) for the lightest chargino mass,
  $ m_{\chi_1^{\pm}} \gsim m_Z/2$,

(ii) $\Gamma (Z \rightarrow \chi_1^0 \chi_1^0) < 0.016 \ {\rm  GeV}, $

(iii) ${\rm Br} (Z \rightarrow \chi^0_i \chi^0_j) \lsim
               1\times 10^{-4}$ except for $(i,j)=(1,1)$.
\newline
The region below the thick line in the $(M_2, \mu)$ plane, which is
excluded by the LEP experiments, does not depend so much on $r$ except for
the region of $M_2 < 50$ (GeV) and $\mu < 0$.
If the mass difference between $\chi_1^{\pm}$ and $\chi_1^0$ is very small,
the electron
(or positron) emitted when the chargino decays to the LSP is too soft to
be identified, and hence it will escape from the direct chargino search.
However,
such an event should be counted as an invisible Z decay. Therefore the bound
$m_{\chi_1^{\pm}} \gsim m_Z/2$ is still valid in this case.

When the Wino is the dominant component of the lightest neutralino,
which is realized for $|r|>1$ and $M_2 \ll |\mu|$, we expect that the
lightest neutralino is highly degenerate in mass with the lightest chargino.
Indeed we can evaluate their masses when $m_W \ll M_2 \ll |\mu|$ and find that
they are degenerate up to the order of $m_W^3$:
\begin{equation}
   M_2 +\frac{m_W^2}{M_2^2-\mu^2} (M_2 +\mu \sin 2 \beta )+
      {\rm higher}.
\end{equation}
We also expect that the masses are quite degenerate when the neutralino is
Higgsino-dominant ($M_2 \gg |\mu|$).
In fig. 2, we  show the mass difference between the lightest
neutralino and the lightest chargino, $\Delta m^{(0)}
= m_{\chi^{\pm}_1}-m_{\chi_1^0}$, at the tree-level for (a)  $r = 5$ and
(b) $r=-5$.  We see that $\chi_1^0$ and $\chi_1^{\pm}$ are
quite degenerate in mass, namely $|\Delta m^{(0)}| < O(1)$ (GeV) in a large
portion of the parameter space.
For $r = 5$, negative $\Delta m^{(0)}$ lies in the region excluded by the LEP
constraints.  Therefore, $\chi_1^{\pm}$ is always heavier than $\chi_1^0$.
On the other hand, for $r = -5$, $\Delta m^{(0)}$
is negative in some part of the parameter space. However, the mass
difference is less than a few tenth GeV.
We have checked that, for other choices of  $r$ ($|r|>1$), the mass degeneracy
generally occurs in the Wino- and Higgsino-dominant regions.

Because the masses are quite degenerate at the tree-level when $|r|>1$,
it is important to consider radiative corrections
to the mass difference  to examine if the LSP is really neutral after
including them.
The calculation of  them in a general parameter region is rather  involved.
Here we will give estimates for two limiting cases, {\it i.e.} $M_2 \ll |\mu|$
and $M_2 \gg |\mu|$.
Let us first define $\Delta m = \Delta m^{(0)}+\Delta m^{(1)}$, where
$\Delta m^{(1)}$ is the one-loop correction to the tree-level mass difference
$\Delta m^{(0)}$.
 For the Wino-LSP case where $M_2 \ll |\mu|$,
the main contribution comes from  gauge boson loops. $\Delta m^{(1)}$ is
calculated to be
\begin{eqnarray}
      \Delta m^{(1)}& =&
      \frac{2g^2 M_2}{(4 \pi)^2}
      \int_0^1 dx (x+1)  \{  \cos^2 \theta_W
        \ln \frac{x^2 M_2^2 + (1-x) m_W^2}{x^2 M_2^2 +(1-x)m_Z^2} \cr
    &  & +  \sin^2 \theta_W
        \ln \frac{x^2 M_2^2 +(1-x) m_W^2}{x^2 M_2^2} \}.
\label{oneloopA}
\end{eqnarray}
{}From eq.~(\ref{oneloopA}), we find that
$\Delta m^{(1)}$ varies from 0.14 (GeV) to 0.17 (GeV) if we vary
 $M_2$= 50 (GeV) to infinity.\footnote{
 For $M_2 \gg m_W$, eq.~(\ref{oneloopA})
reduces to
$ \Delta m^{(1)} =\alpha m_W/2(1+\cos \theta_W)
  + {\cal O} ( m_W^2/M_2 ).
$}
On the other hand, when $M_2 \gg |\mu|$, {\it i.e.} the LSP is almost
a Higgsino,  the gauge boson contribution to $\Delta m^{(1)}$ is
\begin{equation}
\Delta m^{(1)}=\frac{\alpha}{2 \pi} |\mu|
              \int_0^1 dx (x+1) \ln \frac{x^2 \mu^2 +(1-x) m^2_Z}{x^2 \mu^2},
\label{oneloopB}
\end{equation}
which ranges from 0.21 (GeV) to 0.35 (GeV) if we take
$|\mu|$ from 50 (GeV) to infinity.
{}From eqs. (\ref{oneloopA}) and (\ref{oneloopB}),
we can conclude that, in large parameter regions (at least in the Wino-
and Higgsino-dominant regions), the lightest neutralino is indeed
lighter than the lightest chargino. The calculation in the mixed region will
be complicated, which we will not discuss in this paper.

There  is another source which potentially gives a  large radiative
correction to a neutralino mass in the Higgsino LSP region.
 Namely, when the mass-squared mixing
term $m_{\tilde t_{LR}}^2$
between the right
and left stops is large, a top-stop loop will give a radiative correction
to the $\tilde H_2 \tilde H_2$ entry of  the mass
matrix
\begin{eqnarray}
  \Delta &  \sim & \frac{3 h_t^2}{(4 \pi)^2} m_t m_{\tilde t_{LR}}^2 m_S^{-2}
\cr
 &  = & 0.66 ({\rm GeV})
    \times \left( \frac{m_t}{150 ({\rm GeV})} \right)^4
     \left( \frac{m_S}{500 ({\rm GeV})} \right)^{-2}
     \frac{ A}{500 ({\rm GeV})}
     \frac{1}{ \sin^2 \beta},
\end{eqnarray}
where $m_t$ is the top mass, $m_S$ stands for the stop mass scale and
$m_{\tilde t_{LR}}^2 =m_t A$.
This can have both signs and the magnitude can  be as large as 1 GeV or more.

Let us now discuss the effects of relaxing the GUT relation on the possibility
of the LSP dark matter. As we stated above, the LSP is
 a candidate for the dark matter.   It is then an important task
to calculate the cosmic
relic density of the LSPs \cite{Griest,GKT,OliveSrednicki}.
The case of $|r| \ll 1$ has been discussed in detail by Griest and
Roszkowski \cite{GriestRoszkowski}.
They have shown that a light Bino-dominant LSP  with mass $m_{\chi_1^0}
\lsim 10 $ (GeV)
survives the LEP and CDF constraints and that such a light LSP can indeed be a
 candidate  for the dark matter of the Universe.

Here we  discuss the case where $|r|$ exceeds unity
and therefore the LSP can be dominated by the neutral component of the Winos,
$\tilde W_3$.
 If this Wino-dominant LSP is lighter than the W-boson, the annihilation
modes of the neutral Wino pair are
quite similar to those of the Bino pair and hence one might expect that
the light Wino is a  candidate for the dark matter of the Universe.
However this is {\it not} the case.  We have seen that the neutral Wino is
highly degenerate in mass
with its charged counterparts, $\tilde W$.
 At the freeze-out temperature $T \sim m_{\chi_1^0}/20$, the charged Winos  are
as rich as the neutral ones. Then we have to take account of {\it
coannihilation} processes \cite{GriestSeckel,MizutaYamaguchi}
involving superparticles other
than the LSPs, the charged Winos in our case.
Since $\tilde W_3$ and $\tilde W$  can annihilate to a fermion pair through the
 coupling of $\tilde W_3 \gamma^{\mu} \tilde W W^-_{\mu}$,
the relic abundance is greatly
reduced. We use the method of ref. \cite{GriestSeckel}
 to calculate the relic abundance  of the neutralino LSPs,
taking the coannihilations into account.

In fig. 3, we show $\Omega_{\chi}h^2$
where $\Omega_{\chi}$ is the ratio of the mass density of the LSPs to
the critical one to close the Universe and $h$ ($0.4 \leq h \leq 1$) is
the Hubble constant in units of 100 km/s/Mpc.
Here we only consider the case where the LSP is lighter than the W  boson.
We have taken $r=5$ and $\tan \beta=2$. The masses of squarks and sleptons are
assumed to be 1 TeV.
In our numerical calculation, we have
included only quarks and leptons as the final state.
We have used the tree-level mass matrices to obtain the masses of the
neutralinos and charginos. The radiative correction to the mass difference
$\sim 0.2 $ (GeV) does not change our numerical results of the relic
abundance.
When $M_2 \ll |\mu|$ where the LSP is the Wino, the coannihilations between
the neutral and charged Winos explained above make the relic
abundance  of the LSPs very small.
When $M_2 \gg |\mu|$ where the LSP is almost a pure Higgsino,
the relic abundance is small because of the coannihilations among
the Higgsinos \cite{MizutaYamaguchi}, which is quite similar to the GUT case.
When the LSP is a general mixture of the four neutralinos ($M_2 \approx
|\mu|$), the relic abundance is, in general, small because several
annihilation modes  of the LSP pair contribute to reduce it.
 Actually
we can see in fig. 3 the relic abundance is too small for the LSPs to
constitute the dominant component of the energy density of the Universe in
the whole region of the parameter space where the LSP is lighter than
the W-boson. So far we have fixed $\tan \beta=2$ in our calculation. However
we have checked that the results are not sensitive on the choices of $\tan
\beta$.
We have also calculated $\Omega_{\chi} h^2$ for  $r=2$ and $r=1.2$.
In both cases, there exists a tiny region where the LSP is photino-like. It is
interesting to note that this region is within the reach of LEP 200. In this
tiny region, the abundance of the LSPs is sensitive on the sfermion masses
$m_{\tilde f}$. Indeed if we take $m_{\tilde f}=1 $ (TeV) the relic density
exceeds the critical one, while for $m_{\tilde f}=100 $ (GeV) it is very small.
Except for this region, we have seen that the relic density does not depend
on the sfermion masses and it is always much less than the critical density.
If the LSP  is more massive than the W-boson, the LSP pair  annihilates
to the W-pair and the relic abundance is very small.
This annihilation to the W-pair occurs even when the LSP is gaugino-like, the
neutral Wino in this case. This is different from the GUT case: for
a Bino-dominant LSP, this annihilation process is not effective. Thus,
we can conclude  when $|r|>1$  the neutralino LSP is
not cosmologically interesting in most of the parameter region  as far as its
mass is below the TeV scale.\footnote{When the
mass of the LSP increases above the TeV region, $\Omega_{\chi} h^2$ becomes
again of order unity, because there the annihilation cross section is
proportional to $m_{\chi_{1}^0}^{-2}$ and as a result $\Omega_{\chi} h^2
\propto m_{\chi_{1}^0}^{2}$. But such a heavy LSP is not interesting if we
recall that the low-energy SUSY provides a solution of  the naturalness
problem.}

The relaxation of the GUT assumption (\ref{GUTrel}) also affects the mass
spectrum of sfermions, {\it i.e.} squarks and sleptons.  In the $N=1$
supergravity scenario, it is natural to assume that the squarks
 and the sleptons are given a universal mass at some energy scale close to the
Planck one. The actual masses get renormalized when the energy goes down to
the electroweak scale.  Since this RG flow \cite{Inoue}
depends on the gaugino masses, a different mass spectrum will be obtained
from that of the GUT case if we relax the gaugino mass relation imposed by GUT.

In fig. 1, we have shown  that the LSP composition depends strongly on $r$.
 As a result, superparticle searches  become more complicated.
Here, we will consider the  chargino production in electron-positron
collision, $e^+ e^- \rightarrow \chi^+_1 \chi^-_1$ at LEP-200 where the heavier
chargino $\chi_2^{\pm}$ will be too heavy to be accessible in most of the
parameter space. Note that the chargino production does not depend on the
parameter $r$.
 Some  details on the chargino production at LEP-200
 can be found in ref. \cite{BFMM}.
To distinguish the models with different values of $r$, we have to look at the
decay spectrum.  The decays of charginos are numerous, we will restrict
ourselves to the following signature
\begin{equation}
e^+ e^- \rightarrow \chi^+_1 \chi^-_1 ,\  \chi^+_1 \rightarrow \chi^0_1 l^+ \nu
\; .
\end{equation}
We plot the energy and angular distribution for  $ r=$0.5 (solid), 0.25 (dot)
and
1 (dash) in fig. 4 with $\tan \beta = 2$, $M_2 = 100$ (GeV) and $\mu = 250$
(GeV).
 The decay of $\chi_1^+$ involves
the exchanges of a W gauge boson, a sneutrino and a selectron.
The scalar lepton masses are assumed to be 100 GeV.
We find that for larger $r$, the energy of the charged lepton is
softer and the forward-backward asymmetry is smaller.

When $|r|$ exceeds unity, the mass of the
chargino can be nearly degenerate with the LSP mass, and therefore an electron
emitted when the chargino decays is too soft to be identified.
This is actually a difficulty in the chargino search even in
the GUT case, where the mass degeneracy occurs in the Higgsino-dominant
LSP region.
Note, however, in the present case that
the lifetime of the  chargino is roughly given by
\begin{equation}
 \tau = \Gamma (\chi_1^+ \rightarrow \chi_1^0 e^+ \nu )^{-1}
       \sim  10^{-8} {\rm sec} \left( \frac{0.2 {\rm GeV}}{\Delta m} \right)^5,
\end{equation}
where $\Delta m$ denotes the mass difference between the lightest chargino
and the neutralino LSP. If the mass difference is smaller than about
200 (MeV) which can be realized for $|r|>1$ as we have examined,
we have a good chance to trap the chargino  in a detector.

In summary, we have discussed some phenomenological aspects of the SUSY
standard models when the GUT relation on the gaugino masses  is relaxed.
Firstly we have pointed out that the mass and  composition of the lightest
neutralino, which is presumably the LSP, are sensitive
on the parameter $r=M_1/M_2$.
In particular, when $|r|>1$ the neutral Wino can be a dominant
component of the LSP. In this case, the mass of the lightest chargino is
very close to that of the lightest neutralino. We have examined that in
most of the parameter space the neutralino is lighter than the chargino after
including the radiative corrections to the mass difference. Typically the mass
difference is of order 100 (MeV). We have then discussed the cosmic relic
density of the neutralino LSPs. We have shown that, unlike the $|r|<1$ case,
it is  small in most of the parameter space when $|r|>1$ and the neutralino LSP
can not be an interesting
candidate for the dark matter of the Universe. In our discussion, the
coannihilation  processes play an essential role in reducing the relic
density of the LSPs in the Wino- and Higgsino-dominant LSP regions.
  The superparticle searches are also affected by the
parameter $r$. In the chargino search, we may be able to determine $r$ by
looking at the decays of the chargino. When  the mass difference between
$\chi^0_1$ and $\chi^{\pm}_1$ is smaller than $\sim 200$ (MeV), the lifetime
of the chargino becomes long so that it will leave a spectacular
track in a detector.

We would like to acknowledge T. Yanagida for   reading of the manuscript
and useful comments.
Two of us (S.M. and M.Y.) are grateful to K. Inoue, M. Kawasaki,
H. Murayama, X. Tata and
T. Yanagida for helpful discussions. The work of D.N. was supported in part
by the U.S. Department of Energy under Grant No. DE-FG05-85ER-40219.

\newpage


\section*{Figure captions}

\newcounter{Figures}

\begin{list}
{{\bf Fig. \arabic{Figures}.}}
{\usecounter{Figures}}
\item
Contour plots of the mass  and the composition of the LSP in the ($M_2$,
$\mu$) plane when  $r$= (a) 0.1, (b) 0.5 (the GUT case) and (c) 5. Solid lines
show the same mass contours and dot lines show the purity of the Bino
$p(\tilde B)$ for (a) and (b) and of the Wino $p(\tilde W_3)$ for (c) defined
in the text. We have taken $\tan \beta =$2. The region below the thick line is
excluded by the LEP experiments.
\label{fmass}
\item
Contour plots of the mass difference between the lightest chargino and
the lightest neutralino at the tree-level, $\Delta m^{(0)}$, when (a) $r=5$
and (b) $r=-5$. We have taken $\tan \beta =2$.
\label{fdiff}
\item
The relic abundance of the neutralino LSPs when $r=5$ and (a) $\mu>0$
(b) $\mu<0$. We have taken  $\tan \beta=2$, the mass of the pseudoscalar
Higgs boson at
1 TeV and the mass of the sfermions at 1 TeV. The region labelled by LEP is
excluded by the LEP experiments. The LSP is heavier than the W boson in the
region labelled by ``$m_{\chi} > m_W$'',  in which we do not calculate the
relic abundance. Notice that, in the region presented here, $\Omega_{\chi}h^2$
is significantly small.
\label{fabundance}
\item
The energy and angular distributions of the electron emitted by the leptonic
decay of the chargino for $r=0.5$ (solid), 0.25 (dot) and 1 (dash). We have
taken $\tan \beta =2$, $M_2=100$ (GeV) and $\mu=250$ (GeV). The scalar lepton
masses are taken to be 100 (GeV).
\label{fea}
\end{list}


\newpage

\begin{thebibliography}{99}
\setlength{\itemsep}{0cm}
%
%
\bibitem{SU5GUT}
S. Dimopoulos and H. Georgi, Nucl. Phys. {\bf B193} (1981) 150;
N. Sakai, Z. Phys. {\bf C11} (1981) 153.
\bibitem{Inoue}
K. Inoue, A. Kakuto, H. Komatsu and S. Takeshita, Prog. Theor. Phys. {\bf 68}
(1982) 927.
\bibitem{string}
L.E. Ib\'{a}\~{n}ez and D. L\"{u}st, CERN preprint CERN-TH.6380 (1992).
\bibitem{Antoniadis}
I. Antoniadis, J. Ellis, J.S. Hagelin and D.V. Nanopoulos, Phys. Lett. {\bf
B194} (1987) 231.
\bibitem{Drees}
M. Drees, Phys. Lett. {\bf B206} (1988) 265.
\bibitem{DreesTata}
M. Drees and X. Tata, Phys. Rev. {\bf D43} (1991) 2971.
\bibitem{GriestRoszkowski}
K. Griest and L. Roszkowski, CERN  preprint CERN-TH.6181/91 (September 1991).
\bibitem{Yamaguchi}
M. Yamaguchi, North Carolina preprint IFP-426-UNC (February 1992) unpublished.
\bibitem{darkmatter}
H.Goldberg, Phys. Rev. Lett. {\bf 50} (1983) 1419;
L.M. Krauss, Nucl. Phys. {\bf B227} (1983) 556;
J. Ellis, J.S. Hagelin, D.V. Nanopoulos, K.A. Olive and M. Srednicki,
Nucl.Phys. {\bf B238} (1984) 453.
\bibitem{GunionHaber}
J. Gunion and H. Haber, Nucl. Phys. {\bf B272} (1986) 1.
\bibitem{LEP}
ALEPH Collaboration,
D. Decamp {\it et al}, CERN preprint  CERN-PHE/91-149 (1991);
P. Langacker, Pennsylvania preprint UPR-0492T (January 1992).
\bibitem{Griest}
K. Griest, Phys. Rev. {\bf D38} (1988) 2357. E: {\bf D39} (1989) 3802.
\bibitem{GKT}
K. Griest, M. Kamionkowski and M.S. Turner, Phys. Rev. {\bf D41} (1990) 3565.
\bibitem{OliveSrednicki}
K.A. Olive and M. Srednicki, Phys. Lett. {\bf B230} (1989) 78; Nucl. Phys.
{\bf B355 } (1991) 208.
\bibitem{GriestSeckel}
K. Griest and D. Seckel, Phys. Rev. {\bf D43} (1991) 3191.
\bibitem{MizutaYamaguchi}
S. Mizuta and M. Yamaguchi, Tohoku preprint TU-409 (July 1992).
\bibitem{BFMM}
A. Bartl, H. Fraas, W. Majerotto and  B. M\"{o}sslacher,
Z. Phys. {\bf C55} (1992) 257.
\end{thebibliography}
\end{document}